\documentclass[11pt,a4paper]{article}
\usepackage{graphicx}
\usepackage[font=small,labelfont=bf]{caption}
\usepackage{natbib}
\usepackage[ruled,vlined, noend]{algorithm2e}
\usepackage[a4paper, margin=2cm]{geometry}
\usepackage{eurosym}
\usepackage{authblk}
\usepackage{booktabs}
\usepackage[colorlinks=true,citecolor=blue, linkcolor=blue]{hyperref}%
\usepackage{float}
\usepackage[group-separator={,},
            group-four-digits,
            output-decimal-marker={.}]{siunitx}
\usepackage{mathbbol}
\usepackage[draft, textwidth=2cm, textsize=tiny]{todonotes}
\usepackage{amsmath}
\usepackage{amssymb}
\usepackage{url}
\DeclareMathOperator*{\argmax}{arg\,max}
\DeclareMathOperator*{\argmin}{arg\,min}

\begin{document}

\title{Hyper-pool: pooling private trips into high-occupancy transit-like attractive shared rides} 

\author[1]{Rafał Kucharski}
\author[2]{Oded Cats}
\affil[1]{\small Jagiellonian University, Krakow, Poland}
\affil[2]{Delft University of Technology, Delft, the Netherlands}
\maketitle
\begin{abstract}
     We propose Hyper-pool, an analytical, offline, utility-driven ride-pooling algorithm to aggregate individual trip requests into attractive shared rides of high-occupancy. 
     
     We depart from our ride-pooling \texttt{ExMAS} algorithm where single rides are pooled into attractive door-to-door rides and propose two novel demand-side algorithms for further aggregating individual demand towards more compact pooling. First, we generate stop-to-stop rides, with a single pick up and drop off points optimal for all the travellers. Second, we bundle such rides again, resulting with hyper-pooled rides compact enough to resemble public transport operations.
     We propose a bottom-up framework where the pooling degree of identified rides is gradually increased, thereby ensuring attractiveness at subsequent aggregation levels.
     
    
    Our Hyper-pool method outputs the set of attractive pooled rides per service variant for a given travel demand. The algorithms are publicly available and reproducible. It is applicable for real-size demand datasets and opens new opportunities for exploiting the limits of ride-pooling potential. 
    
   In our Amsterdam case-study we managed to pool over 220 travellers into 40 hyper-pooled rides of average occupancy 5.8 pax/veh.

\end{abstract}

\section{Introduction}
\label{intro}

\subsection{Research premise}
Ride-pooling offers a promising alternative in the context of urban mobility, being an intermediate mode between private rides and mass transit. However, its full potential has arguably not been yet realised neither in practice nor in theory. 
In this study, we explore ways to increase the shareability of ride-pooling systems, namely by means of increasing service occupancy and reducing vehicle mileage of ride-pooling systems while remaining strictly attractive for travellers.

We propose a novel kind of pooled rides: \emph{hyper-pool rides}, serving travellers along a sequence of common pick-up and drop off points. To identify attractive hyper-pooled rides, we pool another novel service, \emph{compact stop-to-stop rides}, where travellers ride directly between common pickup and drop-off points. Such attractive compact stop-to-stop rides are, in turn, identified based on \emph{door-to-door pooled rides}, where \emph{private rides} are already pooled together.

\begin{figure}[!htb]
    \begin{center}
      \includegraphics[width=\textwidth]{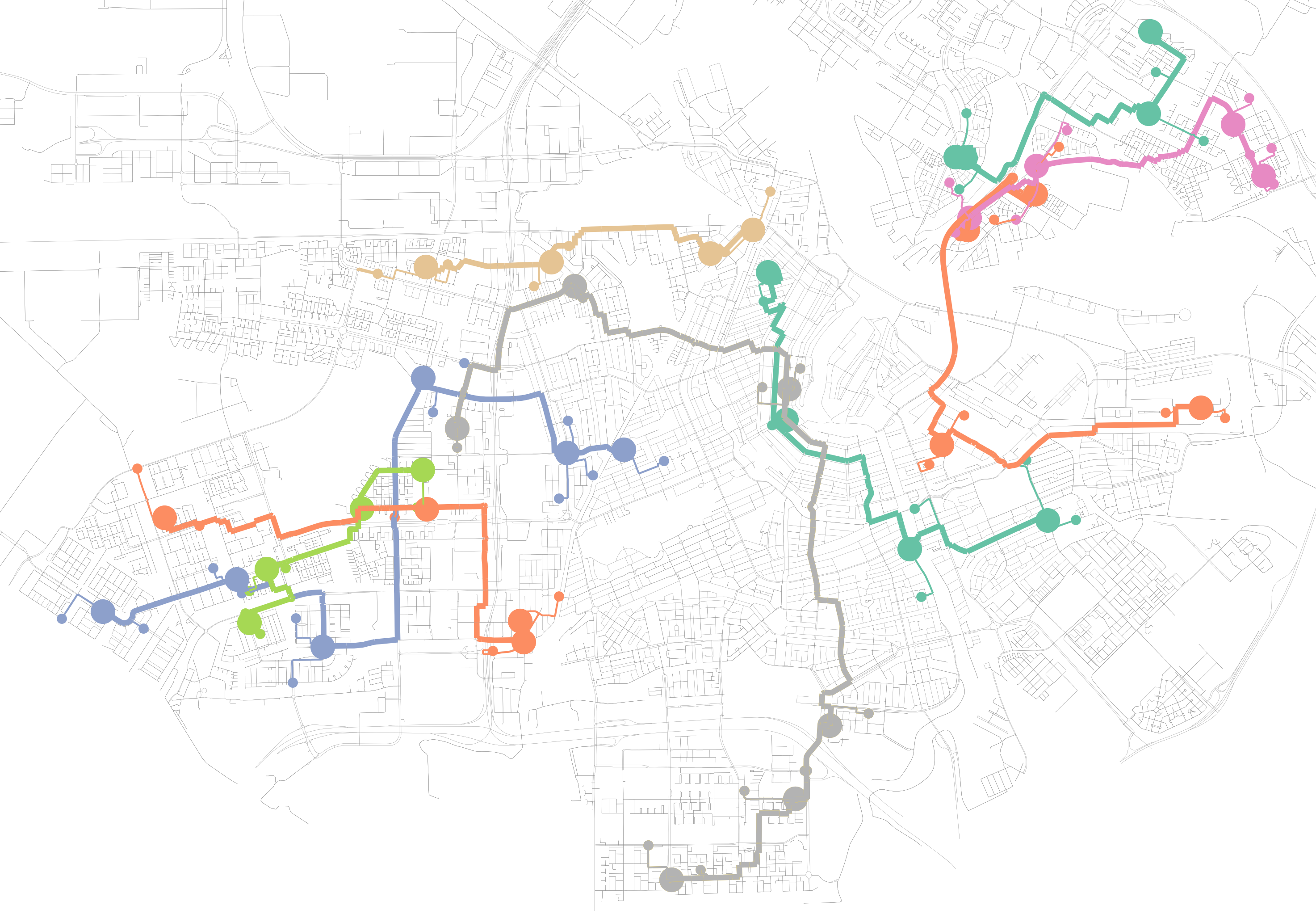}
    \caption{Ten hyper-pooled rides generated for the Amsterdam case study. Each colour denotes a separate multi-stop pooled ride, small dots denote origins and destinations and larger dots denote stops - linked with a walking path marked with lighter lines. The degree of rides (number of travellers) varies from 4 (orange and brown) to 12 (grey and blue) and rides are strictly attractive for all the co-travellers. We can see direct short rides (e.g. brown in the central part) as well as rides spanning through the whole city (grey) and more curved yet still attractive ones (orange). The occupancy typically exceeds four and vehicle hours are reduced five-fold when compared to private rides. The experimental setup and methodology are detailed in further sections.} 
    \label{fig:map1}       
    \end{center}
\end{figure}

Ride-pooling companies primarily offer door-to-door services, picking up and dropping off consecutive travellers along a partially shared route. Unlike mass transit, common pick-up and drop-off points, i.e. stops, are rarely introduced. We introduce pooled rides with common pick-up and drop-off points and demonstrate the benefits of such rides in improving ride-pooling efficiency. In particular, we are interested in the pooled rides which can be attractively composed into a single pick-up and drop-off points, which is crucial for further bundling. These compact stop-to-stop pooled rides are instrumental in introducing the ultimate level of pooling, which already resembles conventional public transport. We bundle multiple stop-to-stop pooled rides into a hyper-pooled ride, where travellers are picked-up at common stop points and the vehicle serves intermediate stops en route. For such rides occupancy may increase substantially, making pooled services efficient and compact enough to resemble public transport operations. Figure \ref{fig:map1} illustrates 10 selected hyper-pooled rides in Amsterdam computed with our proposed algorithm. 

\begin{figure}[!htb]
    \begin{center}
      \includegraphics[width=\textwidth]{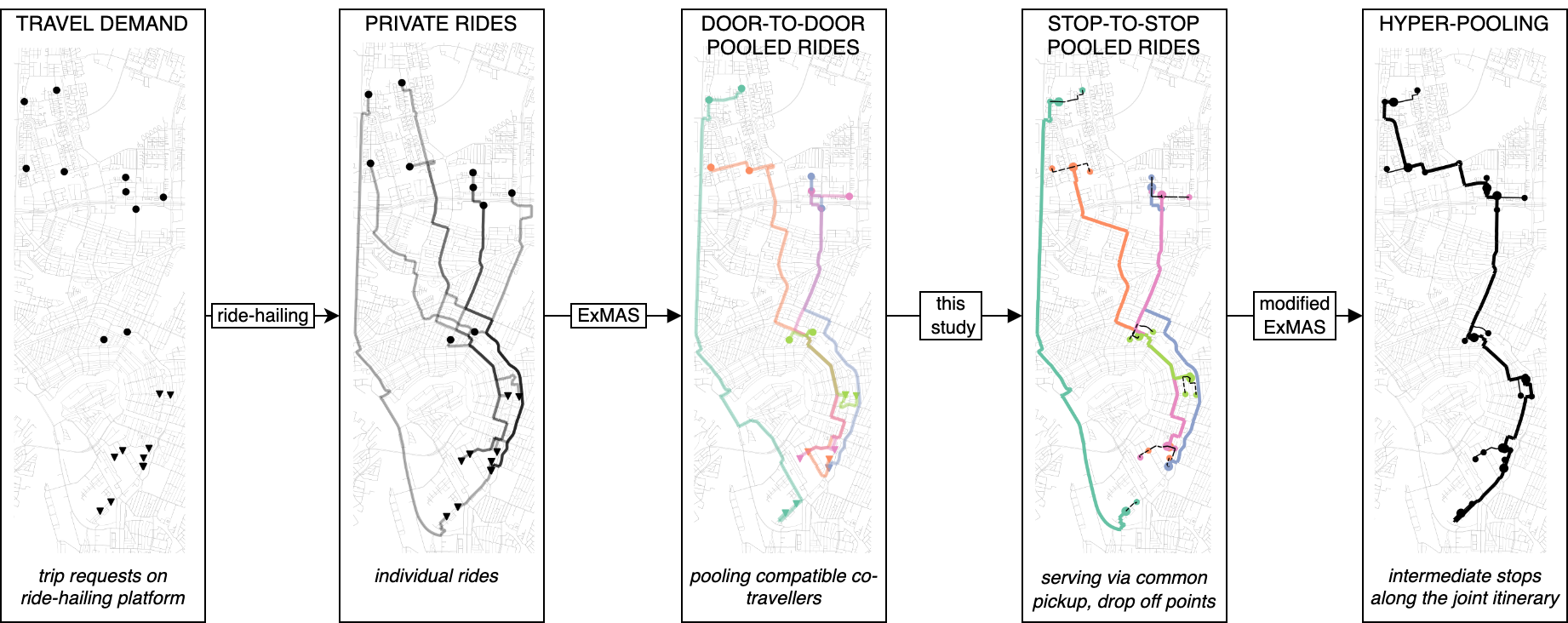}
    \caption{Study conceptual overview. We consider four kinds of rides. Apart from private door-to-door rides, matched with the ExMAS algorithm into door-to-door pooled rides, we introduce two new pooling services: stop-to-stop pooled rides, which are then bundled again into the hyper-pooled rides of high occupancy.}
    \label{fig:scheme}       
    \end{center}
\end{figure}

Hyper-pool is an analytical, offline, utility-driven ride-pooling algorithm. For a given disaggregated travel demand provided in the form of individual trip requests we gradually identify attractive pooled rides of increasing levels of aggregation (as illustrated in fig. \ref{fig:scheme}). 
The main objective is to increase service occupancy levels while remaining attractive for users. The algorithm relies on explicit utility formulas that trade-off any detour, delay and discomfort induced by pooling against the discount offered for sharing the ride.
Since the two newly introduced levels of pooling yield additional discomfort (the first one introduces walking while the second one detours) for them to be attractive for some users this additional discomfort needs to be compensated with increasing levels of discount.
Fortunately, thanks to increasing occupancy the provider may be able to offer lower fares, which can compensate for increasing pooling disutilities, thereby making those sufficiently attractive for users to opt for them.

The solution proposed here revolves around our ExMAS algorithm \citep{exmas}. Door-to-door attractive pooled rides computed with ExMAS are compressed into attractive stop-to-stop rides. Then, identified attractive stop-to-stop pooled rides are themselves pooled once again by the modified ExMAS algorithm. 

Our method is replicable and reproducible on a publicly available source code, parameterizable through the set of design variables (e.g. prices and discounts) and behavioural parameters (value of time, willingness to share, etc.). We tested Hyper-pool for demand levels of up to 8000 trip requests per hour and maximal batches of 2000 travellers. Such levels are large enough for many real-world problems and to induce a critical mass of demand level beyond which attractive hyper-pooled rides emerge.

\subsection{Background}

The potential of ride-pooling to improve urban mobility has triggered a bursting stream of research \citep{agatz2011dynamic, chan2012ridesharing, furuhata2013ridesharing, Tachet2017, bischoff2017city}, which led to an algorithmic breakthrough in the seminal works of \cite{santi2014quantifying} and \cite{alonso2017demand}. 
First, in \cite{santi2014quantifying}, the so-called shareability network was introduced along with a methodology to match travellers in pairs. Further exploited in \cite{alonso2017demand} with the complete algorithm to efficiently match incoming requests with available vehicles by sequentially adding new co-travellers into vehicles with empty seats and adjusting their routes. 
By applying cut-offs on maximal detours and delays the pooled rides are acceptable for travellers and by minimising the service costs (vehicle hours) in the matching the solution becomes optimal for the supply side (platform and/or drivers). 

Recent developments in ride-pooling research proposed a series of improvements to the original solutions, including: 
 algorithmic - to reduce computational complexity \citep{speedup, SIMONETTO2019208, yao2021dynamic};
     heuristic - to narrow the problem search space \citep{liu2020proactive, shah2020neural};
     spatial decomposition - to reduce the problem size \citep{bischoff2018simulation}; or
     demand prediction - to better anticipate future requests \citep{kontou2020reducing, grahn2021improving}.

In parallel, behavioural research has made advancements in better understanding users' so-called willingness to share \citep{kang2021pooled, lavieri2019modeling, alonso2021determinants,lazarus2021pool}. The discomfort of sharing has been quantified and can be now included in classical utility formulas within the random utility modelling framework. Either added as an alternative specific constant (fixed penalty like in \cite{gervzinivc2022potential}) or travel-time multiplier (like in \cite{ li2022pricing} and in this study). Actual detours and delays of pooling were estimated from empirical data \citep{young2020true, zwick2022ride} and the inherent variability of pooling travel and wait time has been incorporated into users' choice models to represent ride-pooling reliability \citep{alonso2020value}.

This allows to reformulate the fixed time-windows of a maximal acceptable detour and delay into utility-based compensation formulas. 
Such, utility-driven approaches, not only put the traveller in the centre but also significantly reduce the search space. Ride-pooling suffers from combinatorial growth of search-space, yet apparently, when rides are initially restricted to attractive ones only, the search space implodes for practical problems and can be searched exhaustively without heuristics.
Such an approach was adopted in our previous study on which we build here \citep{exmas}. 
Similarly to \cite{karaenke2021customer}, who accounts for individual customers’ pooling benefits and compares it to existing provider-centred pooling mechanisms.  
In \cite{ma2022near} the focus is on the sensitivity of the results to user preferences in terms of the level of service (time to be served and excess trip time) and in \cite{de2020ride} willingness to share is formulated as a compensatory cost function at the individual passenger level to explicitly include the utility of pooled travellers.

Ride-pooling can be positioned as an intermediate mode between private ride-hailing and public transport, yet in practice it still remains much closer to the former. While several studies show how ride-hailing may lead to congestion increase \cite{erhardt2019transportation, tarduno2021congestion,bilali2022analytical}, others argue that ride-pooling can potentially become competitive with public transport \cite{hall2018uber, cats2022beyond, erhardt2022transportation}. The critical differences between currently offered pooled services and public transport pertain to fares and travel times. Pooled services are typically more convenient: door-to-door, involving no transfers; yet remain significantly more expensive.

One way to bridge this gap is the introduction of so-called pick-up drop-off points (PUDOs) - which introduce access/egress walk into ride-pooling \citep{stiglic2015benefits}. Such services have been offered for example by Uber (UberExpresPool), MOIA and ViaVan. 
The problem addressed from the operator's perspective focuses on minimising operating costs (vehicle hours and kilometres) within the acceptable access/egress walking distances \cite{fielbaum2021demand}. \cite{GURUMURTHY2022114} experimented with various PUDO spacing and density and reported that average occupancy increased by 20\% and mileage decreased by 27\%. Solutions to this problem are inspired by pickup delivery locations \citep{zhao2018ridesharing} or NP-hard multi-meeting-point route search \citep{li2015optimal}. Conversely, in this study we formulate the problem so that it can be searched exhaustively. 

Research on ride-pooling shares several similarities with on-demand public transport (for a recent review of the latter see~\cite{vansteenwegen2022survey}). The mathematical complexity is similarly challenging and supply-side and demand-side constraints are similar \citep{yoonsimulation}, in both systems a critical mass is needed to take off \citep{Papanikolaou}, yet the strategic planning perspective in the design of on-demand public transport remains more popular than in ride-pooling \citep{narayan2020integrated}. Several recent studies have made initial steps in contributing toward bridging those two communities. \citep{8730311} proposed so-called bus-pooling, they aggregate demand to the cells of predefined OD-matrix and dispatch the vehicle when a certain number of travels are requested. 
An integration between ride-pooling and public transport was proposed for first-mile access to transit \cite{kumar2021algorithm} and indicated a potentiallly significant reduction in vehicle hours travelled. \cite{AUADPEREZ2022103594} combined fixed routes between transit hubs with on-demand shuttles to/from the hubs. In the experiments, costs are 26\% lower than for individual shuttle rides, at the expense of a minimal increase in transit times. A series of experiments with ride-pooling was conducted with MATSim agent-based simulator, recently extended to cover emerging modes, like on-demand and demand-responsive public transport, ride-hailing and ride-pooling \citep{bischoff2018simulation, zwick2020analysis, viergutz2019demand, narayan2021scalability}.

Unfortunately, just when ride-pooling seemed to have gained some critical mass and market shares \citep{wu2022evolution}, COVID-19 impacted it devastatingly \citep{zwick2022ride} with several major players discontinuing their services \citep{vinod2021covid}.

\subsection{Research approach and contribution}

    The full potential of pooled ride-hailing services has not yet been realised. Consequently, we do not know the actual limits of pooled services. The majority of ride-pooling algorithms focus on real-time operations and disregard the attractiveness of the yielded services, implicitly assuming that users are captive. In contrast, demand-oriented studies are limited. 
    
    The central challenge to the ride-pooling problem pertains to its computational complexity - the number of feasible groups of travellers grows combinatorically and even for relatively small problems the search space quickly exceeds reasonable processing capabilities. This is also true for the public transit network design problem, where solutions are limited to heuristic approximations and the discrete search space is not exploitable in practice \citep{vansteenwegen2022survey}. In a previous study, we managed to successfully contain the search space and propose an exact solution for the ride-pooling problem. First, we leverage on the utility and explicitly explored attractive pooling combinations, second, we exploited properties of the so-called shareability graphs and effectively reduced the search space in the hierarchical searches. Here, we follow the same principles to further explore even higher degrees of pooled rides. In our experiments in Amsterdam, we rarely observed door-to-door pooled rides of degree (number of co-travellers) greater than five, which already yielded a search space of size $10^{20}$. While here, we manage to identify rides composed of up to 14 travellers, which would require handling search space of the enormous $10^{57}$ size impossible to search exhaustively\footnote{Calculated as number of 14-element subsets of 2000-element set with every possible sequence of 14 origins and destinations: ${2000 \choose 14}14!14!$}. 
    
    The levels of occupancy and degree like the ones reported here were not identified for attractive pooled rides before. Rides of degree 7 and greater were seldom identified (eg. in \cite{alonso2017demand}) but under a fixed time-window delay, without guarantee of attractiveness. Here, we introduce new services which yield greater slack to compensate for inconveniences of pooling. We extend the hierarchical approach where rides are gradually matched, which drastically reduces the search space. Thanks to this, we identify rides of degree greater than 10, with occupancy greater than 6 (which may allow reducing the base fare six-fold, reaching price levels close to public transit tickets). 
    
    So far, ride-pooling was typically limited to door-to-door services. Stop-to-stop pooled services have not been analysed in an offline, strategic setting. Instead, the focus was typically on real-time vehicle operations and minimising vehicle detours (e.g. like in \cite{fielbaum2021demand}). Consequently, travellers' perspective was only implicitly included via fixed constraints of maximal walking distance, detour and delay. Instead, we explicitly focus on travellers' utility and account for their ability to choose among the various services offered. We identify the stop points at which the utility is maximised for all the co-travellers. Here, we restrict ourselves to the direct stop-to-stop pooled rides only. While typically multiple stops can be identified per ride, we deliberately limit our analysis to rides directly linking a single common pick-up with a single drop-off points. We exploit the compactness of such atomic pooled rides and show how they can be further utilised for the sake of pooling efficiency. Indeed, thanks to their structure they can be now treated as a single trip request. This way we can bundle them again, following the same logic used for composing pooled rides from  private rides. This allows us to explore a hitherto unexploited opportunity.

\section{The Hyper-pool method}

The proposed method utilises an existing algorithm (described in section \ref{ExMAS}), introduces a novel one (in \ref{s2s}) and reformulates the existing one in the new context (in \ref{hp}). Together forming a single pipeline, where trip requests are successively grouped into rides of increasing pooling levels.

\subsection{Input}
The proposed method computes the hyper-pooling for a given demand pattern $\mathbf{Q} = \{Q_1,Q_2, \dots , Q_n \}$, with a single trip $i$ defined through its origin, destination and departure time:
\begin{equation}
    Q_i = ( o_i, d_i, t_i ) \label{trip}
\end{equation}

Origins and destinations are nodes $N$ of the city road network graph ($o_i,d_i \in N$). We assume that the demand is known in advance and remains fixed. The demand stored in such a standard form is available for multiple cities, e.g. New York \citep{dey2021transformation} or Chicago \citep{DEAN2021102944}.

\subsection{Door-to-door ride-pooling with ExMAS.} \label{ExMAS}

We start with computing attractive pooled door-to-door rides with our ExMAS algorithm\footnote{Full methodological details of ExMAS algorithm are available in \cite{exmas} while its practical implementation is publicly available at \url{https://github.com/RafalKucharskiPK/ExMAS}. Here we briefly introduce it, without going into details.}.
ExMAS first identifies the pair-wise shareable trip requests from which it explores the so-called shareability graph to identify pooled rides of gradually increasing degree (number of travellers). 
The feasible pooled rides resulting from ExMAS is the input for the new pooling services proposed here.

In ExMAS, attractive are the rides for which a pooled-ride utility is greater than utility of a private-ride for all the co-travellers. In a door-to-door private (non-shared) ride the utility ($U_p$) is composed of a direct travel time ($t_{p}$) weighted by $\beta^t$ (the value-of-time behavioural parameter) and a distance-based fare ($\lambda_{p} l$):
\begin{equation}
    U_{p} = \beta^t t_{p}- \lambda_{p} l + \varepsilon
    \label{u_p}
\end{equation}
Which, to identify attractive pooling, is compared with the door-to-door pooled ride utility ($U_{d2d}$):
\begin{equation}
    U_{d2d} = \beta^t (\beta^t_{d2d} t_{d2d} + \beta^d d_{d2d}) - \lambda_{d2d}  l + \varepsilon \label{u_d2d}
\end{equation}
Now, the distance-based fare is lower ($\lambda_{d2d}<\lambda_{p}$), which shall at least compensate the longer travel time ($t_{d2d} \geqslant t_{p}$), delay due to pooling ($d_{d2d}>0$) and the discomfort (expressed as $\beta^t_{d2d}>1$ to represent the so-called willingness-to-share)\footnote{Mind that the utility here is in fact a disutility and is negative, due to negative sign in $\lambda$'s and negative values of $\beta^t$. Thus, the computed utility values shall be identified as generalised costs.}.  With the random term in the utility ($\varepsilon$) the model becomes probabilistic, yet here we experiment with the deterministic model ($\varepsilon=0$). To reproduce heterogeneity of travellers, we consider individual value-of-time $\beta^t_i$ which we draw from the normal distribution for each traveller. This way travellers with greater value-of-time will have lower slack of acceptable detours and delays, contrary to those with lower value-of-time .

ExMAS gradually exploits the search space of pooled rides of increasing degree (number of travellers). First it tries to extend identified pairwise shareable trip requests pairs intro triples, then triples into quadruples and so on. ExMAS exhaustively exploits the search space of attractive pooled rides and terminates when no ride can be extended anymore. 

ExMAS outputs the set of attractive rides $\mathbf{R_{d2d}}=\{r_{d2d}\}$, with a generic stop-to-stop ride defined as:
\begin{equation}
    r_{d2d} = ( \mathbf{Q}_r , \mathbf{O}_r ,\mathbf{D}_r, t^p_r),
\end{equation}
, where $\bf{Q}_r$ is a set of served trips, $\bf{O}_r$ and $\bf{D}_r$ are the ordered sequences of served trips' origins and destinations respectively and the $t^p_r$ denotes departure time from the first origin.

Mind that we do not do the matching to assign trip requests to rides (in a so-called  bipartite matching detailed in sect. \ref{match}) at this step, yet we carry all the identified pooled rides to the next pooling levels.

\subsection{Optimal pickup point for stop-to-stop pooled rides} \label{s2s}

We use the output of ExMAS to identify which door-to-door pooled rides can be served from the common pick-up and drop-off points. For each ride, we examine if travellers can attractively walk to a single common pick-up point and arrive at a common drop-off point.
We deliberately restrict ourselves here to the so-called direct stop-to-stop rides and focus on rides which can be served from a single pick-up point to a single drop-off point. The compactness of the obtained solutions is then exploited in the subsequent step of the method.

For a stop-to-stop pooled ride, where travellers walk to a pick-up point, travel directly to the drop-off point and walk to their destinations we introduce the following utility formula:
\begin{equation}
     U_{s2s} = \beta^t (\beta^t_{s2s} t_{s2s} + \beta^d d_{s2s} +  \beta^w w_{s2s}) - \lambda_{s2s}  l +  \varepsilon \label{u_s2s}
 \end{equation}
Similarly to \texttt{ExMAS}, we consider only attractive stop-to-stop pooled rides, i.e. such that $U_{s2s} \geqslant U_{d2d}$, for each co-traveller $i \in \mathbf{Q}_r$.
This means that the reduced price ($\lambda_{s2s}<\lambda_{d2d}$) and presumably shorter travel time ( $t_{s2s,i} \leqslant t_{d2d,i}$) shall at least compensate for the walking discomfort ($\beta^w w_{s2s,i}$). Mind that now we do not assume additional in-vehicle discomfort ($\beta^t_{s2s} \approx \beta^t_{d2d}$) since the ride is anyhow pooled, on the other hand, we assume the walking discomfort is significant ($\beta^w>1$). Moreover, now the travel time is likely to be shorter since the itinerary is reduced to two points only. Those two factors generate an additional slack, thanks to which it is easier to compensate for the negative impact of walking.

We denote such rides $r_{s2s}$ and define them via a set of served travellers ($\mathbf{Q}_r$) and a triplet of origin (pick-up), destination (drop-off) and a departure time:
\begin{equation}
    r_{s2s} = ( \mathbf{Q}_r, o,d,t )
\end{equation}

To identify the optimal stop-to-stop ride, we propose the following exhaustive algorithm\footnote{We did not identify such an exhaustive search as a bottleneck, the search space solutions are small enough to be searched and can be easily parallelised. We avoided analytical solutions, since applying an optimisation program for a non-linear objective function on a multidimensional discrete search space is non-trivial and out of scope for this study.}. To determine the optimal pick-up point, drop-off point and departure time for a given door-to-door pooled ride resulting from ExMAS, we first loop over common pick-up points, then determine the optimal departure time and finally loop over common drop-off points:
\begin{enumerate}
\item First, we identify common access points (accessible from all relevant origins within a fixed threshold $t_w^{max}$), either on a full network graph or only among the predetermined stop locations (e.g. existing bus stops). 
 \item For each of the pickup points reachable from all the origins, we determine the optimal departure time. We first calculate the delay for each traveller as the absolute difference between the desired departure time $t_i$ and the departure time to walk to the pick-up point ($t_w(o_i,o)$) and reach it at the time of common departure $t$:
\begin{equation}
    d_{s2s,i, o , t} = | t_i - t - t_w(o_i,o) | \label{delay}
\end{equation}
From which, we determine the optimal departure time $t$ from stop point $o$ as a minimum of the sum of delays (which are squared to equalise the delay among all the co-travellers):
\begin{equation}
    t = \argmin_{t \in \mathbb{R}} \sum_{i \in r} (d_{s2s, i, o, t})^2 \label{opt_dep}
\end{equation}
This step is independent of the drop-off point $d$ so we may execute it once for each origin and use for all the feasible destinations in the next step.
\item Finally, we explore all the drop-off points from which destinations are reachable within a fixed threshold $t_w^{max}$. 

\end{enumerate}
Now, when we have all the components needed to calculate utility, we may identify the optimal ride. 
We trace how the total utility changes and identify the triplet (pick-up point, departure time and drop-off point) for which it is maximised. Mind that most of the utility components change with the triplet: pick-up point and departure time determine access walk times and delays, drop-off point determines the egress walk times and in-vehicle time changes with both pick-up and drop-off points. Similarly to eq.\ref{opt_dep}, to avoid utility imbalance between co-travellers we apply the so-called logsum formula as follows:
\begin{equation}
    (o,d,t) = \argmax_{o \in N \times t \in \mathcal{R} \times d \in N} \ln \left(  \sum_{i \in r} \exp \left(  U_{s2s,i}(o, t, d)) \right)\right)  \label{argmax}
\end{equation}

While many rides can be transformed into stop-to-stop rides using the aforementioned method, only a small fraction will be attractive for all the travellers. Since we are interested in the attractive poled rides we further consider only such stop-to-stop pooled rides for which we managed to identify such a triplet ($o,d,t$) which makes this service more attractive than door-to-door pooled service for all the co-travellers:
\begin{equation}
\mathbf{R_{s2s}} = \{ (\mathbf{Q}_r, o,d,t) : U_{s2s,i} \geqslant U_{d2d,i} \forall i \in \mathbf{Q}_r \forall r \in r_{d2d}  \},
\end{equation}
which we use as output for the further steps of the algorithm.

In practice, the stop-to-stop pooled rides are identified mostly for the rides of the second degree (two pooled travellers), for the larger number of travellers it is substantially harder to identify common pick-up and drop-off points attractive for more than two co-travellers.
Note that while in the solution we keep both the original door-to-door rides and their stop-to-stop compact forms, in the matching phase (described below) door-to-door rides are likely to be dominated by stop-to-stop rides which typically improve the objective function (vehicle hours).

\subsection{Pooling stop-to-stop rides to the hyper-pooled rides}\label{hp}

Stop-to-stop pooled rides have a similar structure to the private rides, both are defined with a triplet: origin, destination and departure time (compare eqs. \ref{trip} and \ref{argmax}). We can therefore treat them as the input for \texttt{ExMAS} and bundle pooled rides again to what we call hyper-pooled rides. In short, we aim to pool the rides which are already pooled (stop-to-stop rides are composed of several rides, but aggregated to a single pick-up and drop-off point). 

To this end, we express the aggregated utility of a stop-to-stop pooled ride similarly to a private ride in eg.\ref{u_p}, with a formula:
\begin{equation}
    U_{s2s} = \beta^t t_{s2s} - \lambda_{s2s} l + \varepsilon
    \label{u_s2s_aggr}
\end{equation}
Here, the $\lambda_{s2s} l_i$ is the fare paid by the travellers (in stop-to-stop pooled ride the fares are equal among co-travellers) and $t_{s2s}$ is the direct in-vehicle time (from the pick-up stop to the drop-off stop). We keep the already identified optimal pick-up and drop-off stop points and treat them as the origin $o$ and destination $d$ of such a pooled ride respectively. Similarly, we use the optimal departure time computed earlier as the desired departure $t$ for all the already pooled travellers. For the behavioural parameters, we now use the maximal value-of-time $\beta^t$ among the co-travellers and carry it forward as a representative (thanks to such an upper bound we reduce the available slack for pooling, but ensure that the hyper-pooled ride is attractive for the most demanding co-traveller). 

Now, we may identify which of such rides may be pooled again to the hyper-pooled rides $\mathbb{h}$. We use a formula similar to eq.\ref{u_d2d} for door-to-door pooled rides:
\begin{equation}
    U_{\mathbb{h}} = \beta^t (\beta^t_{\mathbb{h}} t_{\mathbb{h}} + \beta^d d_{\mathbb{h}}) - \lambda_{\mathbb{h}}  l + \varepsilon \label{u_ms}
\end{equation}
Similarly to the original \texttt{ExMAS}, the fare is reduced again ($\lambda_{\mathbb{h}}<\lambda_{s2s} $). Now we assume the discomfort due to pooling remains positive ($\beta^t_{\mathbb{h}}>\beta^t_{s2s}$), but is significantly lower than in the original ExMAS (travellers are anyhow pooling).

Similarly to \texttt{ExMAS}, we first identify pairwise shareable rides and match them together. From such pairs, we build a new shareability graph in which we explore pooled rides of gradually increasing degrees. Mind that each stop-to-stop ride is composed of at least two individual trip requests, thus any pair of shareable stop-to-stop rides will pool together at least four individual trip requests.

Some adjustments to the original ExMAS that has to be made when applied for the stop-to-stop pooled rides. One is for overlapping rides. While single trips are pairwise independent, the stop-to-stop rides are composed of individual trips which may be overlapping. For instance, a stop-to-stop ride composed of trips $\{1,2\}$ cannot be pooled with another stop-to-stop ride composed of trips $\{2,3\}$ - since traveller 2 cannot be pooled with herself. To this end we, add a constrain: the pooling candidates need to be mutually exclusive ($\mathbf{Q}_r \cap \mathbf{Q}_r' = \varnothing$).

The second adjustment is to additionally ensure that each ride remains attractive for all the identified hyper-pooled rides, i.e. $U_{\mathbb{h},i}>U_{p,i}$, for each $i \in \mathbf{Q}_r$. Unfortunately, now we test this condition at the aggregated stop-to-stop rides level only, which does not guarantee attractiveness at the individual level (for each single trip request). The inconsistencies at the individual level stem e.g. from various attractiveness of stop-to-stop rides. For some co-travellers, stop-to-stop pooling is very attractive and close to both origin and destination, while for some being at the limit of attractiveness - with long walking time and delay. Similarly, the delay computed with ExMAS needs to be reinterpreted. ExMAS outputs the delay relative to the desired departure time, which for the hyper-pooled rides becomes an optimal departure time computed with (eq.\ref{opt_dep}), thus we need to update the individual delay evaluated with eq. \ref{delay}. For some travellers shifting the departure is positive (moving closer to the desired departure) while for others it is negative.
Nonetheless, we found that using aggregated utility formulas of eq.\ref{u_s2s_aggr} and eq.\ref{u_ms} well approximates the individual utilities faced by co-travellers and only a small share of identified hyper-pool rides do not meet the attractiveness criteria (ca. 5\% travellers) which can be easily post-processed.

With the above minor adjustments, the original ExMAS algorithm successfully pools stop-to-stop rides into hyper-pool rides. The output $\mathbf{R}_{\mathbb{h}}$ is similar to the original ExMAS, but now enriched with info on individual travellers, their access and egress times:
\begin{equation}
    r_{\mathbb{h}} = ( \mathbf{Q}_{s2s} , \mathbf{O}_r ,\mathbf{D}_r, t^p_r),
\end{equation}
$\mathbf{Q}_{s2s}$ denotes a set of stop-to-stop rides pooled together, $\mathbf{O}_r$ and $\mathbf{O}_r$ are the sequences of pickup and drop-off stops, respectively. Access and egress walk times of individual travellers can be obtained from their stop-to-stop rides, while the departure from the first stop $t^p_r$ and the route along the intermediate stops $\mathbf{O}_r$ and $\mathbf{O}_r$ allows calculating their in-vehicle travel times and delays.

\subsection{Trip-ride coverage}
\label{match}

With the previous steps we generated a rich set of feasible rides of various kinds, within which we may now search for the optimal solution. 
Since typically each traveller has multiple options to travel (at the very least a private ride is available and typically multiple alternative rides of various kinds are feasible), we need to find the ride that will be performed by each traveller. 
Our choice set is now composed of rides of four kinds:

\begin{equation}
   \mathbf{R} =  \mathbf{R_p} \cup \mathbf{R_{d2d}} \cup \mathbf{R_{s2s}} \cup \mathbf{R}_{\mathbb{h}}
\end{equation}

To address this, we formulate a so-called trip-ride coverage problem, formulated as an assignment problem, where each trip $i$ is unilaterally matched with a ride  $r\in \mathbf{R} $. We follow the typical formulation \citep{exmas, alonso2017demand} where the assignment is formulated as a problem of identifying a binary vector $x_r$ of length equal to the number of rides $\Vert\mathbf{R}\Vert$, i.e. an assignment variable equal to one if a ride is selected and zero otherwise (eq. \ref{eq:constraint2}). The costs $C_R$ are then expressed as the cost of each ride $c_r$, multiplied by the assignment variable $x_r$ and aggregated for all rides (eq. \ref{Obj}). 
\par
The assignment shall meet the constraint of assigning each trip request to exactly one ride, obtained through the row-wise sum of assignment variable $x_r$ and trip-ride incident matrix $I_{i,r}$. The latter is a binary matrix, in which each entry is one if ride $r$ serves trip $i$ and zero otherwise (eq. \ref{eq:constraint1}).
Eventually, a solution to the problem (eq. \ref{Obj}) is the set of rides $\mathbf{R^*} \subseteq \mathbf{R}$ such that $x_r =1$ $\forall r \in \mathbf{R^*}$. 
We express the matching with the following binary program:
\begin{subequations}
\begin{alignat}{2}
&\!\min        &\qquad& C_R(x_r) = \sum_{r\in \mathbf{R}} c_r x_r \label{Obj}\\
&\text{subject to} &      & \sum_{i \in \mathbf{Q}} I_{i,r} x_r = 1 ,\label{eq:constraint1}\\
&                  &      & x_r \in \lbrace 0,1 \rbrace .\label{eq:constraint2}
\end{alignat}
\end{subequations}

Typically, two perspectives on cost are considered: the one of the service provider (minimising the sum of vehicle hours) and the one of travellers (maximise the sum of utilities $U_{r,i}$). 
Despite taking a strictly demand-oriented stance in this study, we minimise the operator's costs, i.e. vehicle hours. This is since all the pooled rides in the search space are already attractive for all the travellers, thus, when optimising for the service operator, we do not risk the unattractive rides in the solution. Moreover, we are explicitly interested in identifying the potential for high occupancy pooled rides, i.e. those for which vehicle hours are minimal.

Optionally, we may filter the solution $\mathbf{R}$ in eq. \ref{Obj} to consist only of a sub-set of services. For instance, in our experiments, we trace how including new kinds of services improves the system-wide KPIs.

\subsection{Algorithm}

The hyper-pool algorithm solves practical problems in an acceptable time. It is parameterised via custom \texttt{.json} file and can be easily deployed on external computational servers. Open-source \texttt{Python} code is public, allowing for further developments and experiments. The pseudo-code algorithm is presented in algorithm 1 below. It starts with trip requests and gradually searches for attractive rides of four kinds, to finally match travellers to rides by solving the coverage problem.

The hyper-pool algorithm outputs the set of pooled rides. Each trip request (traveller) is assigned to a particular ride. Rides in the solution can be of four kinds: private door-to-door-ride (for those who couldn't be successfully matched with anyone), pooled door-to-door rides, pooled stop-to-stop rides or multi-stop rides. The fare paid by the traveller depends on door-to-door distance and the kind of service.  Other components of travellers' utility: detour, delay and walk time are stored in the solution for further analysis.
\bigskip

\begin{algorithm}[H]
\small
\caption{\texttt{Hyperpool} - pseudo-code for the algorithm}
  \DontPrintSemicolon
  \SetAlgoLined
  \SetKwInOut{Input}{inputs}
  \SetKwInOut{Output}{output}
  \SetKwComment{Comment}{\# }{}
  \SetKw{Break}{break}
  \SetKwProg{HyperPool}{Hyperpool}{}{}

   \HyperPool{}{
     \Input{\\ $\mathbf{Q}$ \Comment*[r]{trip requests}
     $\mathbf{G}$ \Comment*[r]{road network graph}
     $\beta$'s \Comment*[r]{parameters (behavioural, fares, discounts, etc.)}}
     \Output{\\ $\mathbf{R^*}$ \Comment*[r]{pooled rides}}
     
     $\mathbf{R_p} \gets \mathbf{Q}$ \Comment*[r]{Initialise with single rides}
    
     $\mathbf{R_{d2d}} \gets$ ExMAS($\mathbf{R_p},\beta$'s) \Comment*[r]{compute door-to-door pooled rides with \texttt{ExMAS} (sect. \ref{ExMAS})}
     $\mathbf{R_{s2s}} \gets$ Stop-to-Stop($\mathbf{R_{d2d}},\beta$'s) \Comment*[r]{identify stop-to-stop rides with algorithm from sect. \ref{s2s}}
     $\mathbf{R_{\mathbb{h}}} \gets$ ExMAS$'(\mathbf{R_{s2s}},\beta$'s) \Comment*[r]{compute hyper-pooled rides with the modified \texttt{ExMAS} algorithm from sect. \ref{hp}}
     $\mathbf{R} = \mathbf{R_p} \cup \mathbf{R_{d2d}} \cup \mathbf{R_{s2s}} \cup \mathbf{R}_{\mathbb{h}}$ \Comment*[r]{search space of rides of four-kinds}
     $\mathbf{R^*} \subseteq \mathbf{R}$ \Comment*[r]{match travellers to rides by solving the coverage problem in sect. \ref{match}}
    }
\SetAlgoLined
\KwResult{$\mathbf{R^*}$}

\end{algorithm}
\bigskip

Despite the radical reduction of search space, the hyper-pool, (similarly to ExMAS) explodes under specific settings (beyond some demand levels and discount rates). Nonetheless, we managed to compute 10-minute batches with up to 2000 trip requests in less than two hours on a standard laptop (MacBook Pro 2021 M1 chip, 16GB RAM), which seems to be sufficient for offline strategic analyses of the most practical ride-pooling problems. Furthermore, the stop-to-stop rides algorithm can be parallelised (each pooled ride in a separate thread), which can significantly reduce the computation times. The bottleneck of the current implementation seems to be in the second run of ExMAS when stop-to-stop rides are bundled into hyper-pooled rides. Since at this step we do not penalise sharing with an additional multiplier ($\beta^t_{\mathbb{h}}$) and the discount is significant, we often end up with a huge number of identified hyper-pooled rides (over a million in our half-hour batch, which covered only 250 out of 2000 individual trip requests) - this may corrupt the bipartite-matching  (eq. \ref{eq:constraint1} evaluated now on a matrix $I_{i,r}$ with 2000 rows and over a million columns).

\subsection{The solution and KPIs}

The set of pooled rides resulting from hyper-pool provides the set of indicators to measure the service performance (KPI). In particular, we observe the:
\begin{itemize}
    \item vehicle hours, i.e. total time spent by vehicles to serve the demand (empty kilometres, like deadheading and repositioning, are not included here)
    \item passenger utility, i.e. total (dis)utility of travellers, both total as well as decomposed into: 
    \begin{itemize}
        \item trip fare (direct monetary costs)
        \item in-vehicle travel time
        \item walk time (to access and egress to/from stop points)
    \end{itemize}
    \item fare-efficiency, i.e. vehicle travel time divided with the total fare. Expressed in \euro~ per vehicle kilometre and used as a proxy of service efficiency for the provider.
    \item and the occupancy:
    \begin{equation}
        ( \sum_{i \in \mathbf{Q} } t_{p} ) / ( \sum_{r \in \mathbf{R^{*}} } t_{r} ) 
    \end{equation}     \label{occ}

The occupancy is, in our opinion, the most meaningful one. We have a particular definition here, which we argue for. Typically, in public transit the occupancy is calculated for the actual passenger-hours, yet here we use the shortest-path passenger-hours, i.e. the total passenger-hours for the private rides $\left( \sum_{i \in \mathbf{Q} } t_{p} \right)$.  In the denominator, we take the actual vehicle hours of the selected rides $\left(\sum_{r \in \mathbf{R^{*}} } t_{r} \right) $.  Such a measure is sensitive to detours (which are included in the denominator only), it directly traces the compactness of pooling.
\end{itemize}

\section{Results}
We demonstrate the results on a case study of Amsterdam (The Netherlands). We begin with the illustrative example of 10 travellers pooled together into the attractive multi-stop shared ride. Then, we report the results from the case of 2000 Amsterdam trips requested in half an hour. 

\paragraph{Experimental setting}
We sample trip requests from the Albatross dataset \citep{arentze2004learning} using Amsterdam PM peak trips longer than 2km. We treat this demand as fixed and do not consider opt-outs, the travellers can only choose among offered ride-hailing and ride-pooling services.
We assume value-of-time to be normally distributed around 12\euro/h with standard deviation $\sigma=1.5$\euro/km to represent natural travellers' heterogeneity.
We start with the actual Amsterdam fare for private ride-hailing rides of 1.5\euro/km and assume the following discounts: 25\% for door-to-door ride-pooling, 66\% for stop-to-stop and 75\% for multi-stop pooling, i.e. the fares $\lambda$'s of 1.5, 1.11, 0.5 and 0.375 \euro/km respectively (PT fare in Amsterdam is ca. 0.3 \euro/km, only 25\% more expensive than the proposed hyper-pool rides). We penalise sharing with $\beta^t_{d2d}=1.3$, and use 1.5 multipliers for walking ($\beta^d$) and 1.2 for waiting ($\beta^d$). Common catchment areas are searched in the radius of 7.5 minutes walk time around origins and destinations ($t_w^{max}$). We assume fixed and network-wide flat speeds: 1.5 m/s for walking and  8 m/s (28.8 km/h) for rides (which can be substituted with space- and time-varying speeds). We use the detailed OSM graph of Amsterdam to get travel times and distances \citep{boeing2017osmnx}. To obtain attributes for public transport alternatives we use the actual GTFS public transport timetable and Open Trip Planner query engine (following the method described in \cite{cats2022beyond}).

Initially, we simulated relatively short, 10-minute batches, which, despite higher demand levels,  were too short to accommodate longer hyper-pool trips. Thus we simulated a longer, 30-minute batch with 2000 trip requests (4000 trips/hour). Such setting was found at one hand realistic, with actual values for the Amsterdam setting (value-of-time, ride-hailing price, average speed, etc.) and on another allowed to reproduce hyper-pooling with attractive shared rides of high occupancy attractive for travellers and plausible for providers under mild subsidises.

\subsection{An illustrative example}

\begin{figure}[H]
\begin{center}
  \includegraphics[width=\textwidth]{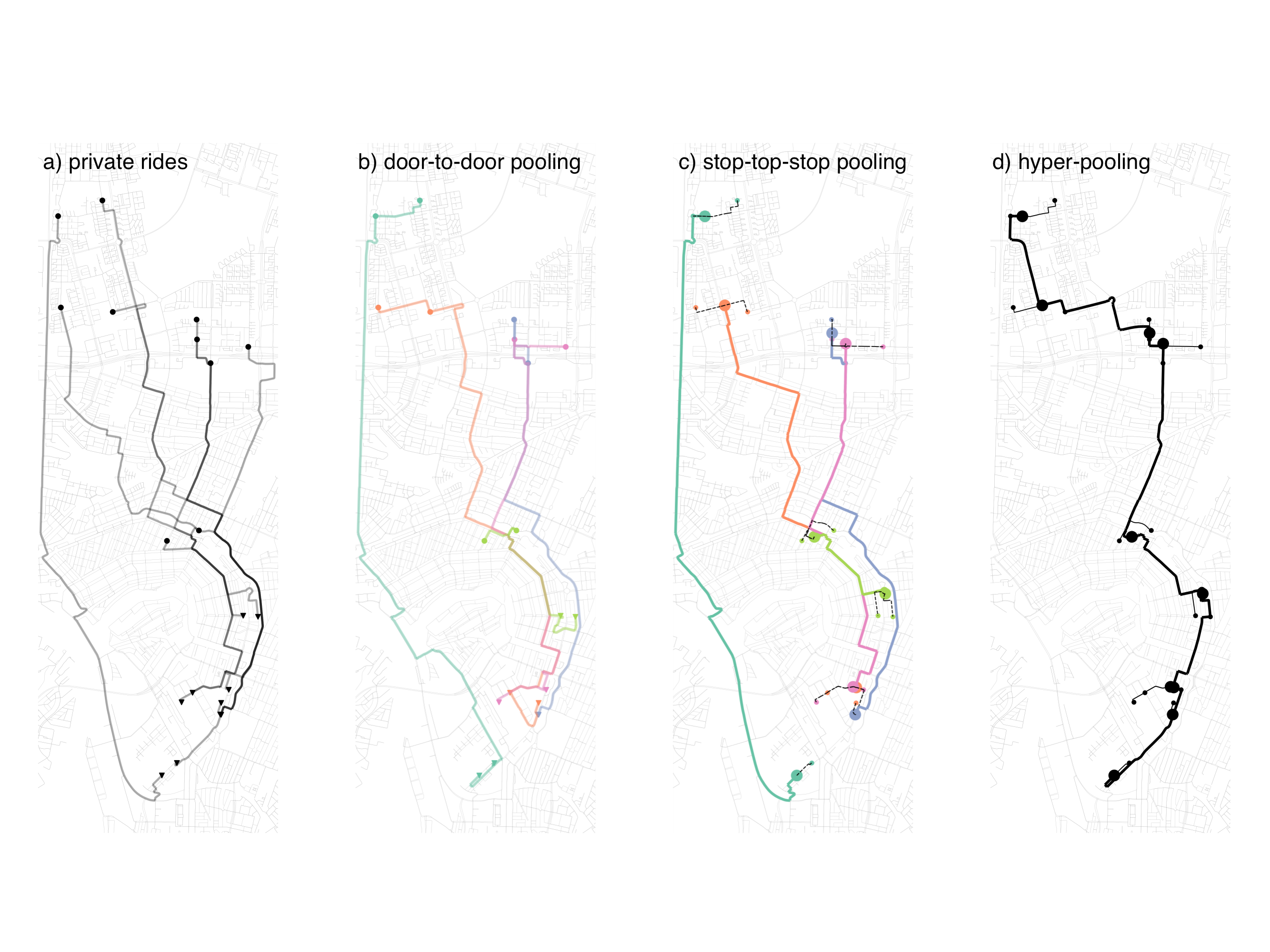}
\caption{Set of 10 private trip requests (a) bundled with our algorithm to a single, attractive hyper-pooled ride (d). First, via the set of door-to-door shared rides (b), then compressed into stop-to-stop shared rides with walking marked with dashed lines (c) and finally into a single hyper-pooled ride (d). Numerical details are presented on table \ref{tab:1}.}
\label{fig:map}       
\end{center}
\end{figure}

First, we illustrate 10 trip requests from Amsterdam PM peak travel demand which the proposed algorithm managed to bundle into a single hyper-pooled ride. Figure \ref{fig:map} shows consecutive pooling levels from  private rides (a) via consecutive steps of pooling: door-to-door pooled rides (b), stop-to-stop rides (c) and finally bundled into a single hyper-pooled ride on (d).
Nine requests of a total 65km length were served with a 9.6km multi-stop ride. If served with private rides it generates 2.25 vehicle-hours, while it can be now served with only 0.44 (as detailed in table. \ref{tab:1}). Total walking time is 78 minutes and total in-vehicle time is 157 minutes. Nonetheless, the total disutility for each traveller in this ride is lower than for the private ride and for some travellers it is even better than the public transport (PT) alternative. The total in-vehicle travel times remain twice shorter than PT and the total walking time 30\% lower. Total disutility drops from 127 to 95\euro, which is lower than PT (112\euro). Notably, the total collected fare drops from \euro 97 for private rides to 24 for a multi-stop ride. To remain attractive, the ride-hailing operator would need to reduce his revenues four times (from 97 to 24\euro), which need to be compensated by cost reductions (vehicle hours) or a subsidy. Generating this particular subset of trips and ordering both origins and destinations without our algorithm would require searching in the enormous search space of $3.86\times 10^{40}$ combinations.

\begin{table}[H]
\caption{An illustrative example of a set of 10 travellers  bundled into a single hyper-pooled ride ($\mathbb{h}$) (depicted in fig. \ref{fig:map}). Rows denote consecutive travellers. In columns, we report (dis)utilities, travel times, walk times and fares paid at the four levels of pooling: private ($p$), door-to-door ($d2d$), stop-to-stop ($s2s$), hyper-pool ($\mathbb{h}$); and public transport (PT) alternative.}
\resizebox{\columnwidth}{!}{
\begin{tabular}{l|SSSSS|SSSSS|SSSSS|SSSSS}
\toprule
{} & \multicolumn{5}{l}{utility} & \multicolumn{5}{l}{travel time [minutes]} & \multicolumn{5}{l}{walk time} & \multicolumn{5}{l}{fare [\euro]} \\
id &       {p} &    {d2d} &    {s2s} &     { $\mathbb{h}$} &     {PT} &            {p} &    {d2d} &    {s2s} &     { $\mathbb{h}$} &   {PT} &           {p} &    {d2d} &    {s2s} &   { $\mathbb{h}$} &    {PT} &   {p} &    {d2d} &    {s2s} &     { $\mathbb{h}$} &     {PT}  \\
\midrule
776  & -10.26 &  -9.78 &  -9.72 &  -7.42 &  -6.46 &   12.23 &   12.23 &   12.23 &   15.63 &    31.78 &     0.0 &     0.0 &    6.82 &    6.82 &    10.23 &    8.82 &    6.48 &    6.48 &    2.20 &     2.36 \\
811  & -19.33 & -17.87 & -19.15 & -14.12 & -15.09 &   20.65 &   20.65 &   20.65 &   26.88 &    50.83 &     0.0 &     0.0 &   10.35 &   10.35 &     9.73 &   14.87 &   10.93 &   10.93 &    3.72 &     3.59 \\
1106 & -18.97 & -18.44 & -18.42 & -12.49 & -14.52 &   20.12 &   20.12 &   20.12 &   26.88 &    39.75 &     0.0 &     0.0 &    1.82 &    1.82 &     6.78 &   14.49 &   10.66 &   10.66 &    3.62 &     2.93 \\
1162 & -14.09 & -13.98 & -14.06 & -13.17 & -15.69 &   13.93 &   13.93 &   12.43 &   15.63 &    33.67 &     0.0 &     0.0 &    2.23 &    2.23 &     6.92 &   10.04 &    7.38 &    3.41 &    2.51 &     2.61 \\
1385 & -15.53 & -15.48 & -14.66 & -10.29 & -12.53 &   16.82 &   16.82 &   16.82 &   18.90 &    43.87 &     0.0 &     0.0 &   12.77 &   12.77 &    12.43 &   12.11 &    8.91 &    8.91 &    3.03 &     2.98 \\
1401 & -16.06 & -15.76 & -15.98 & -11.71 & -16.66 &   16.27 &   16.27 &   16.27 &   18.90 &    47.48 &     0.0 &     0.0 &    9.68 &    9.68 &    18.67 &   11.72 &    8.62 &    8.62 &    2.93 &     2.81 \\
1470 & -13.52 & -13.00 & -12.98 & -10.82 & -11.60 &   14.05 &   14.05 &   12.43 &   13.68 &    35.22 &     0.0 &     0.0 &   13.15 &   13.15 &    14.38 &   10.12 &    7.44 &    3.44 &    2.53 &     2.31 \\
1729 & -12.82 & -12.42 & -12.42 &  -9.36 & -13.17 &   12.82 &   12.82 &   12.82 &   13.68 &    35.23 &     0.0 &     0.0 &    4.37 &    4.37 &    13.93 &    9.23 &    6.79 &    6.79 &    2.31 &     2.34 \\
1865 &  -3.08 &  -2.92 &  -2.89 &  -3.05 &  -3.98 &    3.63 &    3.63 &    3.48 &    3.68 &    18.38 &     0.0 &     0.0 &    9.92 &    9.92 &    12.65 &    2.62 &    1.93 &    0.89 &    0.66 &     1.36 \\
1995 &  -3.84 &  -3.33 &  -3.60 &  -2.71 &  -2.92 &    4.72 &    4.72 &    3.48 &    3.68 &    15.28 &     0.0 &     0.0 &    7.65 &    7.65 &     9.55 &    3.40 &    2.50 &    1.16 &    0.85 &     1.36 \\ \midrule
total & -127.5 & -122.98 & -123.88 & -95.64 & -112.62 &  135.24 &  135.24 &  130.73 &  157.54 &   351.49 &     0.0 &     0.0 &   78.76 &   78.76 &   115.27 &   97.42 &   71.64 &   61.29 &   24.36 &    24.65 \\
\bottomrule
\end{tabular}}
\label{tab:1}   
\end{table}

\subsection{Case study: Half an hour in Amsterdam.}

We now turn to reporting results from the case of 2000 trips requested in a 30-minute batch (4000 trips/hour). We design the experiment to show the impact of consecutive levels of pooling, i.e. we start with private rides, add door-to-door rides with ExMAS, identify stop-to-stop rides and finally hyper-pooled rides. At each of the introduced steps, we solve the matching and see how the newly added rides impact the solution. 

The resulting rides' compositions obtained at respective pooling levels are shown in table \ref{tab:2}. Each newly introduced level of pooling attracts a significant portion of travellers. Door-to-door poling is attractive for over 1300 travellers and 171 of them can be successfully compacted into stop-to-stop rides. Eventually, over 225 travellers are successfully matched into attractive hyper-pooled rides. Notably, introducing new levels of pooling does not reduce the number of private rides in the solution, which remains stable and around 30\% of travellers were not matched with any other co-traveller. Thus the impact of the newly introduced pooled rides is mainly on further compacting travellers who are already pooling. The number of feasible pooled rides (last column) identified with ExMAS was over 150 000, which only slightly increased with identified stop-to-stop rides, to finally explode into over a million hyper-pooled feasible rides.

Most importantly, the demand of 225 travellers was successfully supplied with only 40 hyper-pooled rides composed on average of 5.6 travellers. If every one of 225 travellers would travel alone, they would require 52.5 vehicle-hours - in the hyper-pooled multi-stop rides it is reduced six-fold to 9 vehicle hours only, allowing to achieve the average occupancy of 5.8. If we treat vehicle-hours reduction as a proxy for cost reductions, this might allow to discount the fare to a level comparable with public transport. The latter is on average five times cheaper than ride-hailing.

\begin{table}[H]
    \caption{Rides composition obtained when pooling 2000 trip requests at various pooling levels. Each row introduces a new level of pooling and columns denote the number of travellers assigned to rides of respective kind. The last column denotes the number of attractive rides in respective solutions.}
    \label{tab:2}       
    \begin{center}
    \resizebox{0.8\columnwidth}{!}{
    \begin{tabular}{ll|rrrr|S}
    \toprule
    {} & {} &  private &  door-to-door &  stop-to-stop &  hyper-pooled &  {number of } \\
    solution with   &         &          &                      &                      &               &     {feasible rides}       \\
    \midrule
    private & $p$             &     2000 &                    0 &                    0 &             0 &       2000 \\
    door-to-door pooled & $p \cup d2d$ &      655 &                 1345 &                    0 &             0 &     159702 \\
    stop-to-stop pooled & $p \cup d2d \cup s2s $ &      645 &                 1184 &                  171 &             0 &     160239 \\
    hyper-pooled  & $p \cup d2d \cup s2s \cup \mathbb{h} $ &      651 &                 1065 &                   59 &           225 &    1009855 \\
    \bottomrule
    \end{tabular}}
       
    \end{center}
\end{table}

Introducing new levels of pooling improves all the major KPIs (tab. \ref{tab:3}). Vehicle hours are reduced, traveller's costs (disutility) are lower, in-vehicle hours are reduced and occupancy rates increase. Vehicle hours are reduced from 274 hours for private rides, to 171 for the solution with all four kinds of pooling services. The total passenger generalised travel times, i.e. increase in disutility, is reduced from 7.56\euro to 6.65\euro per passenger. In-vehicle hours are lowest for private rides, then increase for door-to-door pooling, drop when stop-to-stop rides are introduced and drop again when hyper-pooling is introduced. This reduction is achieved by inducing very short walking times. The mean fare collected by the service provider drops from almost 5.92\euro to 4.02\euro per passenger. The average occupancy of pooling (computed with eq. \ref{occ}) remains at the stable level of around 1.5 and increases only slightly when new pooling kinds are introduced (from 1.53 in door-to-door to 1.60 in hyper-pooled). The greatest efficiency (fares per vehicle hours) is obtained for door-to-door pooling, introducing hyper-pooling is not directly attractive for the service provider, whose efficiency drops from 50.8 to 47 \euro~per vehicle hour, yet remains at a significantly higher level when compared to the private ride-hailing solution (47 versus 43\euro/veh-km). 

\begin{table}[H]
    \caption{Ride-pooling KPIs per traveller obtained for case of 30-minute batch with 2000 trips in Amsterdam at four consecutive levels of pooling.}
    \label{tab:3}       
    \resizebox{\columnwidth}{!}{
    \begin{tabular}{l|SSrrrrrr}
    \toprule
    {} &  {vehicle} &  {(dis)utility} &  {pax in-vehicle} &  {walk time}  &  {\#rides} &  {fare} &  {fares per veh-hour} &  {Average occupancy} \\
    solution            &      {hours}          &     {\euro}                        &  {hours}                           &             &                 &             &                      &            \\
    \midrule
    private             &         0.137 &                      -7.56 &                      0.137 &        0.00 &            2000 &       5.92 &                43.24 &       1.00 \\
    door-to-door pooled &         0.090 &                      -7.31 &                      0.181 &        0.00 &          159702 &        4.55 &                50.81 &       1.53 \\
    stop-to-stop pooled &         0.088 &                      -7.18 &                      0.169 &       0.01 &          160239 &        4.35 &                49.43 &       1.56 \\
    hyper-pooled   &             0.086 &                      -6.65 &                      0.146 &       0.02 &         1009855 &        4.02 &                47.03 &       1.60 \\
    \bottomrule
    \end{tabular}}
\end{table}

\begin{figure}[H]
\begin{center}
  \includegraphics[width=\textwidth]{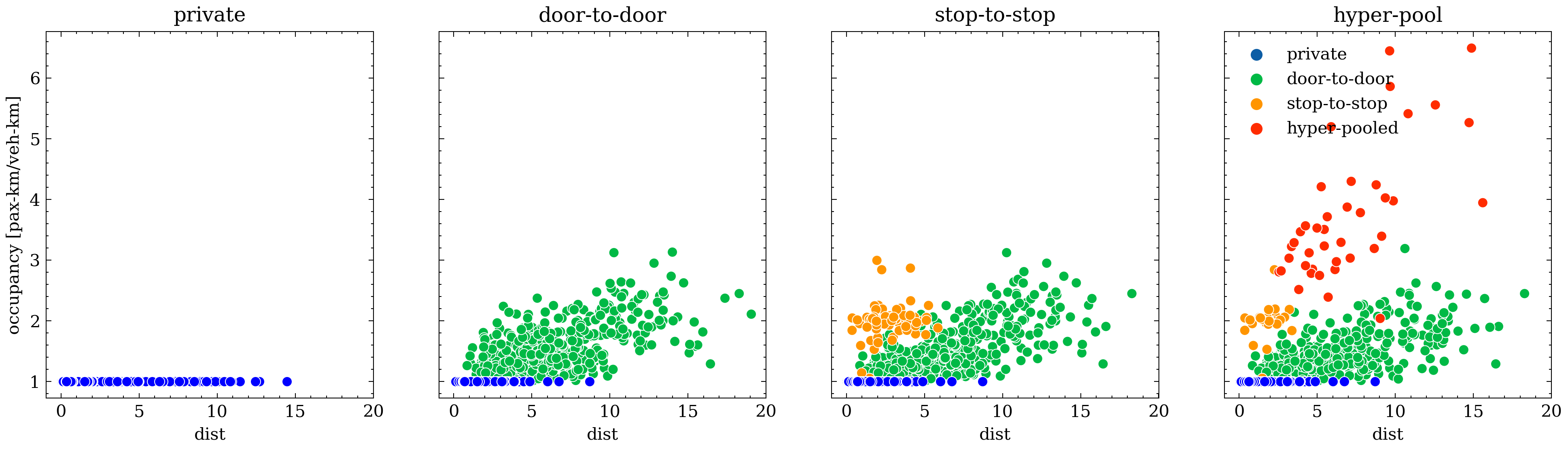}
\caption{Occupancy (y-axis) of rides obtained at respective levels of ride pooling scattered against ride length (x-axis). Each panel denotes a solution where new levels of pooling are introduced, dots represent individual rides and colours denote the pooling service kind. Private rides (a) have equal occupancy. Door-to-door services of ExMAS (b) dominate for longer trips but are also present for the shorter trips. Stop-to-stop trips (c) are identified mainly for the shorter trips and do not induce significantly greater occupancy. However, when they are bundled again into hyper-pooled rides not only the rides are longer, but the occupancy reaches significantly higher levels.}
\label{fig:c1}       
\end{center}
\end{figure}

\begin{figure}[H]
\begin{center}
  \includegraphics[width=0.8\textwidth]{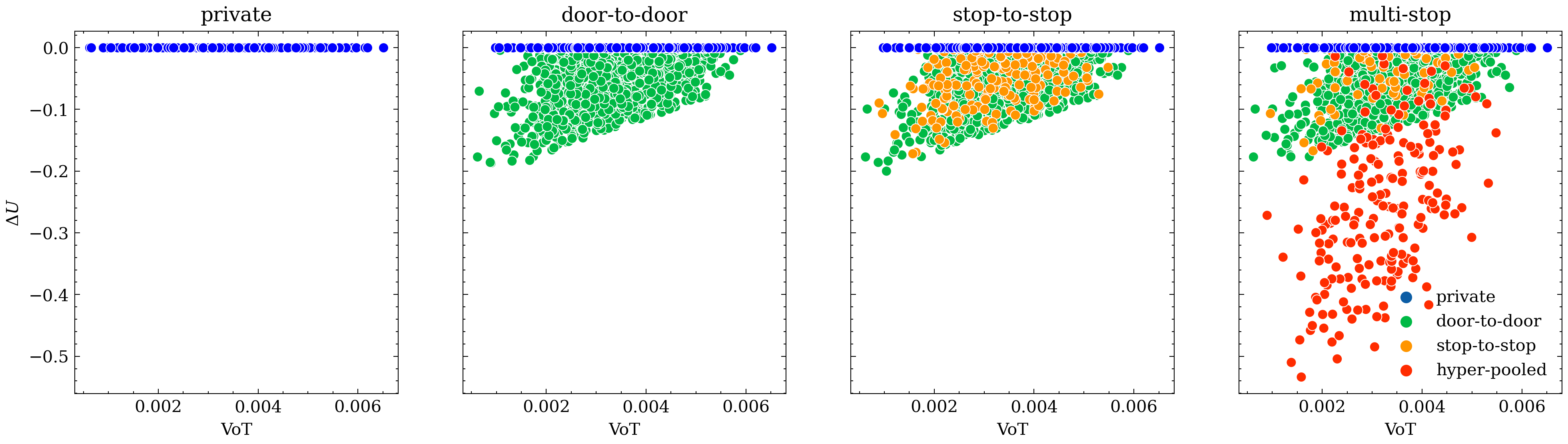}
\caption{Relative attractiveness (disutility reduction $\Delta U$ on y-axis as related to the private ride) obtained at various levels of pooling (consecutive panels) scattered against the value of time of respective travellers (x-axis). Each dot denotes now a single traveller and is coloured accordingly to the type of pooling service to which the traveller was assigned to.}
\label{fig:c2}       
\end{center}
\end{figure}

\begin{figure}[H]
\begin{center}
  \includegraphics[width=0.8\textwidth]{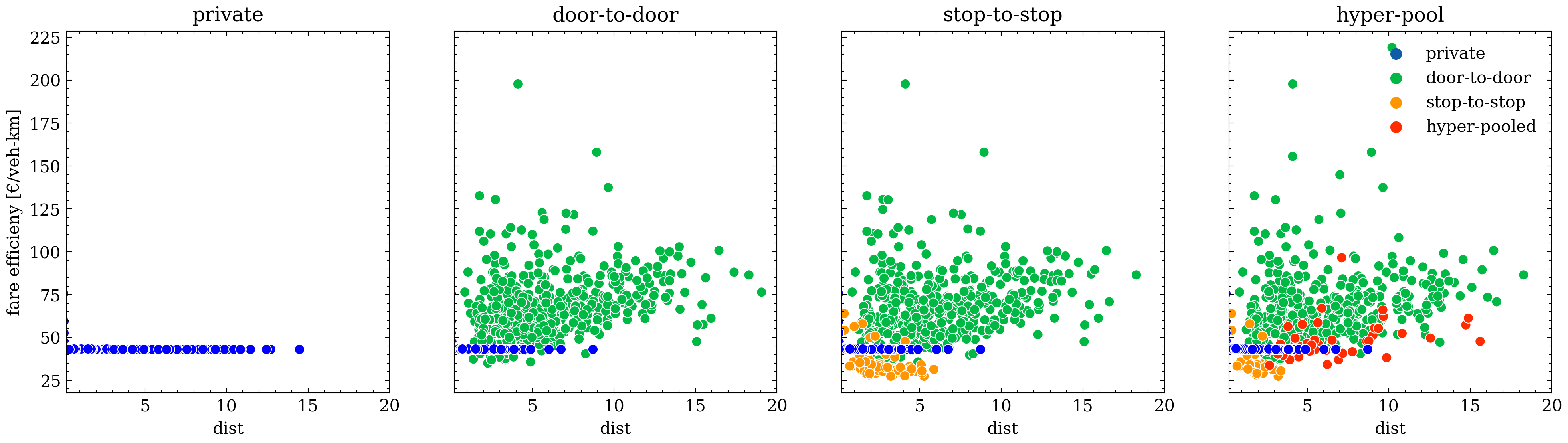}
\caption{Trip efficiency (fare per vehicle-kilometre on y-axis) varying with ride length (x-axis) for various ride types (colours) and pooling levels (panels). Each dot represents a single ride.}
\label{fig:c3}       
\end{center}
\end{figure}

More detailed analysis of individual trips is presented in figs. \ref{fig:c1}, \ref{fig:c2}, and \ref{fig:c3}. There we present, in the consecutive panels, how introducing new levels of pooling impacts the solution and distinguish rides/travellers levels of pooling with distinct colours. 

First, we show occupancy for rides of various lengths in fig. \ref{fig:c1}. We see the massive impact of hyper-pooled rides on the last panel, reaching occupancies levels not observed in the other pooling services. The stop-to-stop rides (introduced in panel c) are not the main contributor to system efficiency, they are rather mean to an end, i.e. being instrumental to propose the hyper-pooled rides composed of them. The compact stop-to-stop rides have occupancy comparable to door-to-door rides but are typically shorter. We can see a similar trend twice on the picture, first: longer private trips (blue) are more often pooled into door-to-door rides (green) - compare panel a with b; second: longer door-to-door rides (orange) are more often pooled into hyper-pooled rides (red) - compare panel c with d.

Figure \ref{fig:c2} shows how attractive are pooled rides of various kinds for travellers. We show how the travellers' utility improves relative to a private ride. Here each dot denotes one of 2000 travellers and x-axis distinguishes travellers' heterogeneous value-of-time ($\beta^t$).
We see a strong trend and lower bound of attractiveness for door-to-door pooling (b) - the relative gains of pooling are greater for travellers with lower value-of-time. The stop-to-stop pooling (c) impact on travellers' utility is negligible. Hyper-pooling, however, breaks this trend and sharply reduces (dis)utilities, which can now be halved. The main trend can be still seen, but is not so evident anymore - effective hyper-pooled rides are found for travellers with high value-of-time as well. Similarly to the impact on occupancy, the stop-to-stop rides are not attractive per se, they do not pose benefits for the travellers (despite a significant discount: 1.11\euro/km for a door-to-door ride is reduced in this experiment to 0.5\euro/km). This discount is just enough to compensate for walking (which has a high perceived discomfort), yet the true benefits come when hyper-pooled services are offered. Up to a two-fold reduction in disutility is observed among the travellers opting for a pooled ride.

Finally, in fig. \ref{fig:c3}, we focus on the perspective of the service provider. Obviously, under the fixed demand, offering discounts always yields lower total revenues, though ride-pooling allows reducing the costs as well. To analyse it, we trace total fares paid and divide it by vehicle-hours, which we treat as a proxy of operating costs. While, due to flat per-kilometre fare and fixed network speeds, private rides (panel a) provide a flat of 44 \euro /veh-hour (x-axis) for door-to-door rides efficiency varies significantly (panel b). For some small number of rides door-to-door pooled rides are less efficient, yet for the majority effectiveness increases and reaches up to 200 \euro /veh-hour (e.g. when five highly compatible rides are pooled together). Introducing stop-to-stop rides (panel c) is typically ineffective and the discount is not compensated with reduced mileage. Many stop-to-stop rides are twice less attractive for the service provider than private rides. Again, similarly to the two previous figures, this trend is broken with hyper-pooled rides, the majority of which is more beneficial for the service provider than private rides. This suggests that under such settings, subsidies for the service provider may not be necessary. These disaggregated results may be compared with fare per veh-hour reported in the table. \ref{tab:3}: introducing hyper-pooling reduces effectiveness compared to the door-to-door rides (from 50 to 47), yet it remains more cost-effective when compared to private rides only (47 to 43).

\section{Discussion and conclusions}

Our proposed hyper-pooling method yields occupancy levels not observed before for attractive ride-pooling. Up to 14 travellers pooled together in rides where the discomfort induced by pooling is compensated by reduced fares. It offers a great potential for travellers (whose disutility of travelling can be reduced), for policymakers (who contribute towards sustainability goals with increased occupancy) and for service providers (for whom the pooling cost-effectiveness remains greater than for private ride-hailing). The main contribution pertains to the generation of hyper-pooled rides, they break the current limits of the classical ride-pooling. The intermediate service of stop-to-stop rides is instrumental, yet their own attraction value is found to be limited. 

Despite only 11\% of the travellers have been successfully served by hyper-pooled rides in our experiment, the impact of such rides on the pooling is significant. Obviously, the critical mass in the demand levels is greater than for classical pooling, i.e. more travellers are needed to identify attractive rides of high degree. Yet already 4000 trip requests per hour allow over 10\% of travellers to travel with hyper-pooled rides. Notably, we did not consider in our study induced demand and modal shifts. If we assume that more attractive pooling will induce the demand for the services and some car travellers will opt for this, this can further increase efficiency. For several of our travellers, the hyper-pooled rides were even more attractive than the public transport alternative (in Amsterdam, where the quality of PT is considered high).

We set the experiment to reach the fare levels similar to the costs of public transit (0.3\euro/km for PT and 0.375\euro/km for hyper-pooling), expecting such service to be highly attractive for travellers, yet not cost-efficient for the service provider. Surprisingly, the significant mileage reductions and high occupancy effectively compensated for the service provider. Without taking into account the cost of larger vehicles, the cost-effectiveness of hyper-pooled rides exceeds the one of private ride-hailing. 

Similarly, we devised our experiment to account for travellers' heterogeneous value-of-time, expecting to observe that hyper-pooled rides are composed of travellers with lower values of time. Surprisingly, our algorithm managed to effectively compensate travellers with greater values of time and find hyper-pooled rides also for them. 

\paragraph{Caveats and limitations}

Despite the radical reduction in search space, our method is not exhaustive and relies on assumptions that prevent us from claiming that our solution is globally optimal. Since we gradually increase the discount with the newly introduced pooling levels, it yields inconsistencies as we hierarchically search rides of increasing pooling levels. In the original ExMAS we assumed a consistent incentive for sharing, which made our method exact. Here we increase the incentive (reduced fare) for already pooled travellers, thus it is not guaranteed that we exploit all the feasible groups for the multi-stop pooled rides. Notwithstanding, we manage to identify highly valuable solutions meeting a series of practical constraints.

\paragraph{Public subsidises}
Whether discounts offered for hyper-pool services shall remain attractive also for the system provider, remains an open question. Traditionally ride-pooling services are considered a commercial product offered by profit-oriented companies, yet as soon as they reach occupancy levels comparable to transit we may start treating them as public services. They can thus be subsidised for sake of sustainability and reaching policy goals based on their added-value to urban accessibility or in contrast, ban them when they cannibalise public transport(see \cite{cats2022beyond}) . In our analysis we limit our effectiveness constraints to the demand-side only, allowing for unprofitable rides as well. 

\paragraph{Feet operations} In this study we deliberately take the traveller perspective and treat the fleet operations as exogenous.  We are interested in the potential of a given demand to be pooled and assume that any solution will be served by a fleet of sufficient size. In other words, we assume that the fleet operator will provide enough vehicles to serve the pooled rides without additional delays beyond those caused by the compromise among the co-travellers to minimise joint departure delays.
Similarly, our vehicle-hours (mileage) are only a rough approximation of the actual costs. First, we do not consider dead-heading and repositioning and focus on the distance travelled with a passenger on-board only. Second, we do not make a distinction between different vehicle types. Rides composed of more than three travellers require special kinds of vehicles, resembling minibuses - presumably more expensive to operate which may be taken into consideration in future research.

\paragraph{Future directions} We identify the following additional directions for further research.First, in our procedure stops are set as optimal for travellers, while results often show that a minor shift may significantly reduce the vehicle route and avoid detours (as e.g. in  \cite{fielbaum2021demand}). This can improve the solution and further reduce mileage. 
Second, the hyper-pooled rides may happen to be attractive for other travellers, currently assigned to the private rides. This can be easily checked: for each private (non-shared) ride in the solution we may search for a hyper-pool ride that a traveller finds attractive (has greater utility than the private ride). We do not expect this to have a large effect, since if the traveller cannot be matched in the solution with anyone, he is unlikely to be compatible with a hyper-pool ride composed of at least four other travellers. 

We believe the proposed approach is not only beneficial for improving the efficiency of ride-pooling but may also inspire the development of novel solutions to the transit network design problems, which to date remain open.  While resulting hyper-pool rides are not yet public transport lines, they may be used as spatial patterns driving the network design solutions towards a new optima. Broader experiments with larger travel demand sets may reveal repetitive patterns and clustering methods may spatially and temporary aggregate hyper-pooled rides into potential service lines.

\paragraph{Acknowledgements}

This research was supported by the CriticalMaaS project (Grant Number 804469), which is financed by the European Research Council and Amsterdam Institute for Advanced Metropolitan Solutions. This research was funded by National Science Centre in Poland program OPUS 19 (Grant Number 2020/37/B/HS4/01847).

\bibliographystyle{spbasic}      

\bibliography{bib}   

\begin{thebibliography}{55}
\providecommand{\natexlab}[1]{#1}
\providecommand{\url}[1]{{#1}}
\providecommand{\urlprefix}{URL }
\expandafter\ifx\csname urlstyle\endcsname\relax
  \providecommand{\doi}[1]{DOI~\discretionary{}{}{}#1}\else
  \providecommand{\doi}{DOI~\discretionary{}{}{}\begingroup
  \urlstyle{rm}\Url}\fi
\providecommand{\eprint}[2][]{\url{#2}}

\bibitem[{Agatz et~al(2011)Agatz, Erera, Savelsbergh, and
  Wang}]{agatz2011dynamic}
Agatz N, Erera AL, Savelsbergh MW, Wang X (2011) Dynamic ride-sharing: A
  simulation study in metro atlanta. Procedia-Social and Behavioral Sciences
  17:532--550

\bibitem[{Alonso-Gonz{\'a}lez et~al(2020)Alonso-Gonz{\'a}lez, {van Oort}, Cats,
  Hoogendoorn-Lanser, and Hoogendoorn}]{alonso2020value}
Alonso-Gonz{\'a}lez MJ, {van Oort} N, Cats O, Hoogendoorn-Lanser S, Hoogendoorn
  S (2020) Value of time and reliability for urban pooled on-demand services.
  Transportation Research Part C: Emerging Technologies 115:102,621

\bibitem[{Alonso-Gonz{\'a}lez et~al(2021)Alonso-Gonz{\'a}lez, Cats, van Oort,
  Hoogendoorn-Lanser, and Hoogendoorn}]{alonso2021determinants}
Alonso-Gonz{\'a}lez MJ, Cats O, van Oort N, Hoogendoorn-Lanser S, Hoogendoorn S
  (2021) What are the determinants of the willingness to share rides in pooled
  on-demand services? Transportation 48(4):1733--1765

\bibitem[{Alonso-Mora et~al(2017)Alonso-Mora, Samaranayake, Wallar, Frazzoli,
  and Rus}]{alonso2017demand}
Alonso-Mora J, Samaranayake S, Wallar A, Frazzoli E, Rus D (2017) On-demand
  high-capacity ride-sharing via dynamic trip-vehicle assignment. Proceedings
  of the National Academy of Sciences 114(3):462--467

\bibitem[{Arentze and Timmermans(2004)}]{arentze2004learning}
Arentze TA, Timmermans HJ (2004) A learning-based transportation oriented
  simulation system. Transportation Research Part B: Methodological
  38(7):613--633

\bibitem[{Auad-Perez and {Van Hentenryck}(2022)}]{AUADPEREZ2022103594}
Auad-Perez R, {Van Hentenryck} P (2022) Ridesharing and fleet sizing for
  on-demand multimodal transit systems. Transportation Research Part C:
  Emerging Technologies 138:103,594,
  \doi{https://doi.org/10.1016/j.trc.2022.103594},
  \urlprefix\url{https://www.sciencedirect.com/science/article/pii/S0968090X22000407}

\bibitem[{Bilali et~al(2022)Bilali, Fastenrath, and
  Bogenberger}]{bilali2022analytical}
Bilali A, Fastenrath U, Bogenberger K (2022) Analytical model to estimate ride
  pooling traffic impacts by using the macroscopic fundamental diagram.
  Transportation Research Record p 03611981211064892

\bibitem[{Bischoff et~al(2017)Bischoff, Maciejewski, and
  Nagel}]{bischoff2017city}
Bischoff J, Maciejewski M, Nagel K (2017) City-wide shared taxis: A simulation
  study in berlin. In: 2017 IEEE 20th international conference on intelligent
  transportation systems (ITSC), IEEE, pp 275--280

\bibitem[{Bischoff et~al(2018)Bischoff, Kaddoura, Maciejewski, and
  Nagel}]{bischoff2018simulation}
Bischoff J, Kaddoura I, Maciejewski M, Nagel K (2018) Simulation-based
  optimization of service areas for pooled ride-hailing operators. Procedia
  Computer Science 130:816--823

\bibitem[{Boeing(2017)}]{boeing2017osmnx}
Boeing G (2017) Osmnx: New methods for acquiring, constructing, analyzing, and
  visualizing complex street networks. Computers, Environment and Urban Systems
  65:126--139

\bibitem[{Cats et~al(2022)Cats, Kucharski, Danda, and Yap}]{cats2022beyond}
Cats O, Kucharski R, Danda SR, Yap M (2022) Beyond the dichotomy: How
  ride-hailing competes with and complements public transport. Plos one
  17(1):e0262,496

\bibitem[{Chan and Shaheen(2012)}]{chan2012ridesharing}
Chan ND, Shaheen SA (2012) Ridesharing in north america: Past, present, and
  future. Transport reviews 32(1):93--112

\bibitem[{Dean and Kockelman(2021)}]{DEAN2021102944}
Dean MD, Kockelman KM (2021) Spatial variation in shared ride-hail trip demand
  and factors contributing to sharing: Lessons from chicago. Journal of
  Transport Geography 91:102,944,
  \doi{https://doi.org/10.1016/j.jtrangeo.2020.102944},
  \urlprefix\url{https://www.sciencedirect.com/science/article/pii/S0966692320310218}

\bibitem[{Dey et~al(2021)Dey, Tirtha, Eluru, and
  Konduri}]{dey2021transformation}
Dey BK, Tirtha SD, Eluru N, Konduri KC (2021) Transformation of ridehailing in
  new york city: A quantitative assessment. Transportation Research Part C:
  Emerging Technologies 129:103,235

\bibitem[{Engelhardt et~al(2020)Engelhardt, Dandl, and Bogenberger}]{speedup}
Engelhardt R, Dandl F, Bogenberger K (2020) Speed-up heuristic for an on-demand
  ride-pooling algorithm. \doi{10.48550/ARXIV.2007.14877},
  \urlprefix\url{https://arxiv.org/abs/2007.14877}

\bibitem[{Erhardt et~al(2019)Erhardt, Roy, Cooper, Sana, Chen, and
  Castiglione}]{erhardt2019transportation}
Erhardt GD, Roy S, Cooper D, Sana B, Chen M, Castiglione J (2019) Do
  transportation network companies decrease or increase congestion? Science
  advances 5(5):eaau2670

\bibitem[{Erhardt et~al(2022)Erhardt, Mucci, Cooper, Sana, Chen, and
  Castiglione}]{erhardt2022transportation}
Erhardt GD, Mucci RA, Cooper D, Sana B, Chen M, Castiglione J (2022) Do
  transportation network companies increase or decrease transit ridership?
  empirical evidence from san francisco. Transportation 49(2):313--342

\bibitem[{Fielbaum et~al(2021)Fielbaum, Bai, and
  Alonso-Mora}]{fielbaum2021demand}
Fielbaum A, Bai X, Alonso-Mora J (2021) On-demand ridesharing with optimized
  pick-up and drop-off walking locations. Transportation research part C:
  emerging technologies 126:103,061

\bibitem[{Furuhata et~al(2013)Furuhata, Dessouky, Ord{\'o}{\~n}ez, Brunet,
  Wang, and Koenig}]{furuhata2013ridesharing}
Furuhata M, Dessouky M, Ord{\'o}{\~n}ez F, Brunet ME, Wang X, Koenig S (2013)
  Ridesharing: The state-of-the-art and future directions. Transportation
  Research Part B: Methodological 57:28--46

\bibitem[{Ger{\v{z}}ini{\v{c}} et~al(2022)Ger{\v{z}}ini{\v{c}}, van Oort,
  Hoogendoorn-Lanser, Cats, and Hoogendoorn}]{gervzinivc2022potential}
Ger{\v{z}}ini{\v{c}} N, van Oort N, Hoogendoorn-Lanser S, Cats O, Hoogendoorn S
  (2022) Potential of on-demand services for urban travel. Transportation pp
  1--33

\bibitem[{Grahn et~al(2021)Grahn, Qian, and Hendrickson}]{grahn2021improving}
Grahn R, Qian S, Hendrickson C (2021) Improving the performance of first-and
  last-mile mobility services through transit coordination, real-time demand
  prediction, advanced reservations, and trip prioritization. Transportation
  Research Part C: Emerging Technologies 133:103,430

\bibitem[{Gurumurthy and Kockelman(2022)}]{GURUMURTHY2022114}
Gurumurthy KM, Kockelman KM (2022) Dynamic ride-sharing impacts of greater trip
  demand and aggregation at stops in shared autonomous vehicle systems.
  Transportation Research Part A: Policy and Practice 160:114--125,
  \doi{https://doi.org/10.1016/j.tra.2022.03.032},
  \urlprefix\url{https://www.sciencedirect.com/science/article/pii/S0965856422000878}

\bibitem[{Hall et~al(2018)Hall, Palsson, and Price}]{hall2018uber}
Hall JD, Palsson C, Price J (2018) Is uber a substitute or complement for
  public transit? Journal of urban economics 108:36--50

\bibitem[{Kang et~al(2021)Kang, Mondal, Bhat, and Bhat}]{kang2021pooled}
Kang S, Mondal A, Bhat AC, Bhat CR (2021) Pooled versus private ride-hailing: A
  joint revealed and stated preference analysis recognizing psycho-social
  factors. Transportation Research Part C: Emerging Technologies 124:102,906

\bibitem[{Karaenke et~al(2021)Karaenke, Schiffer, and
  Waldherr}]{karaenke2021customer}
Karaenke P, Schiffer M, Waldherr S (2021) The customer is always right:
  Customer-centered pooling for ride-hailing systems. arXiv preprint
  arXiv:210701161

\bibitem[{Kontou et~al(2020)Kontou, Garikapati, and Hou}]{kontou2020reducing}
Kontou E, Garikapati V, Hou Y (2020) Reducing ridesourcing empty vehicle travel
  with future travel demand prediction. Transportation Research Part C:
  Emerging Technologies 121:102,826

\bibitem[{Kucharski and Cats(2020)}]{exmas}
Kucharski R, Cats O (2020) Exact matching of attractive shared rides (exmas)
  for system-wide strategic evaluations. Transportation Research Part B:
  Methodological 139:285--310, \doi{https://doi.org/10.1016/j.trb.2020.06.006},
  \urlprefix\url{https://www.sciencedirect.com/science/article/pii/S0191261520303465}

\bibitem[{Kumar and Khani(2021)}]{kumar2021algorithm}
Kumar P, Khani A (2021) An algorithm for integrating peer-to-peer ridesharing
  and schedule-based transit system for first mile/last mile access.
  Transportation Research Part C: Emerging Technologies 122:102,891

\bibitem[{Lavieri and Bhat(2019)}]{lavieri2019modeling}
Lavieri PS, Bhat CR (2019) Modeling individuals’ willingness to share trips
  with strangers in an autonomous vehicle future. Transportation research part
  A: policy and practice 124:242--261

\bibitem[{Lazarus et~al(2021)Lazarus, Caicedo, Bayen, and
  Shaheen}]{lazarus2021pool}
Lazarus JR, Caicedo JD, Bayen AM, Shaheen SA (2021) To pool or not to pool?
  understanding opportunities, challenges, and equity considerations to
  expanding the market for pooling. Transportation Research Part A: Policy and
  Practice 148:199--222

\bibitem[{Li et~al(2022)Li, Jiang, and Lo}]{li2022pricing}
Li M, Jiang G, Lo HK (2022) Pricing strategy of ride-sourcing services under
  travel time variability. Transportation Research Part E: Logistics and
  Transportation Review 159:102,631

\bibitem[{Li et~al(2015)Li, Qin, Yu, and Mao}]{li2015optimal}
Li RH, Qin L, Yu JX, Mao R (2015) Optimal multi-meeting-point route search.
  IEEE Transactions on Knowledge and Data Engineering 28(3):770--784

\bibitem[{Liu et~al(2019)Liu, Zhang, and Yang}]{8730311}
Liu K, Zhang J, Yang Q (2019) Bus pooling: A large-scale bus ridesharing
  service. IEEE Access 7:74,248--74,262, \doi{10.1109/ACCESS.2019.2920756}

\bibitem[{Liu and Samaranayake(2020)}]{liu2020proactive}
Liu Y, Samaranayake S (2020) Proactive rebalancing and speed-up techniques for
  on-demand high capacity ridesourcing services. IEEE Transactions on
  Intelligent Transportation Systems

\bibitem[{Ma and Koutsopoulos(2022)}]{ma2022near}
Ma Z, Koutsopoulos HN (2022) Near-on-demand mobility. the benefits of user
  flexibility for ride-pooling services. Transportation Research Part C:
  Emerging Technologies 135:103,530

\bibitem[{Narayan et~al(2020)Narayan, Cats, van Oort, and
  Hoogendoorn}]{narayan2020integrated}
Narayan J, Cats O, van Oort N, Hoogendoorn S (2020) Integrated route choice and
  assignment model for fixed and flexible public transport systems.
  Transportation Research Part C: Emerging Technologies 115:102,631

\bibitem[{Narayan et~al(2021)Narayan, Cats, van Oort, and
  Hoogendoorn}]{narayan2021scalability}
Narayan J, Cats O, van Oort N, Hoogendoorn SP (2021) On the scalability of
  private and pooled on-demand services for urban mobility in amsterdam.
  Transportation Planning and Technology pp 1--17

\bibitem[{Papanikolaou and Basbas(2021)}]{Papanikolaou}
Papanikolaou A, Basbas S (2021) Analytical models for comparing demand
  responsive transport with bus services in low demand interurban areas.
  Transportation Letters 13(4):255--262, \doi{10.1080/19427867.2020.1716474},
  \urlprefix\url{https://doi.org/10.1080/19427867.2020.1716474},
  \eprint{https://doi.org/10.1080/19427867.2020.1716474}

\bibitem[{de~Ruijter et~al(2020)de~Ruijter, Cats, Alonso-Mora, and
  Hoogendoorn}]{de2020ride}
de~Ruijter A, Cats O, Alonso-Mora J, Hoogendoorn S (2020) Ride-sharing
  efficiency and level of service under alternative demand, behavioral and
  pricing settings. In: Transportation Research Board 2020 Annual Meeting

\bibitem[{Santi et~al(2014)Santi, Resta, Szell, Sobolevsky, Strogatz, and
  Ratti}]{santi2014quantifying}
Santi P, Resta G, Szell M, Sobolevsky S, Strogatz SH, Ratti C (2014)
  Quantifying the benefits of vehicle pooling with shareability networks.
  Proceedings of the National Academy of Sciences 111(37):13,290--13,294

\bibitem[{Shah et~al(2020)Shah, Lowalekar, and Varakantham}]{shah2020neural}
Shah S, Lowalekar M, Varakantham P (2020) Neural approximate dynamic
  programming for on-demand ride-pooling. In: Proceedings of the AAAI
  Conference on Artificial Intelligence, vol~34, pp 507--515

\bibitem[{Simonetto et~al(2019)Simonetto, Monteil, and
  Gambella}]{SIMONETTO2019208}
Simonetto A, Monteil J, Gambella C (2019) Real-time city-scale ridesharing via
  linear assignment problems. Transportation Research Part C: Emerging
  Technologies 101:208 -- 232, \doi{https://doi.org/10.1016/j.trc.2019.01.019},
  \urlprefix\url{http://www.sciencedirect.com/science/article/pii/S0968090X18302882}

\bibitem[{Stiglic et~al(2015)Stiglic, Agatz, Savelsbergh, and
  Gradisar}]{stiglic2015benefits}
Stiglic M, Agatz N, Savelsbergh M, Gradisar M (2015) The benefits of meeting
  points in ride-sharing systems. Transportation Research Part B:
  Methodological 82:36--53

\bibitem[{Tachet et~al(2017)Tachet, Sagarra, Santi, Resta, Szell, Strogatz, and
  Ratti}]{Tachet2017}
Tachet R, Sagarra O, Santi P, Resta G, Szell M, Strogatz SH, Ratti C (2017)
  {Scaling Law of Urban Ride Sharing}. Nature Publishing Group pp 1--6,
  \doi{10.1038/srep42868}, \urlprefix\url{http://dx.doi.org/10.1038/srep42868}

\bibitem[{Tarduno(2021)}]{tarduno2021congestion}
Tarduno M (2021) The congestion costs of uber and lyft. Journal of Urban
  Economics 122:103,318

\bibitem[{Vansteenwegen et~al(2022)Vansteenwegen, Melis, Akta{\c{s}},
  Montenegro, Vieira, and S{\"o}rensen}]{vansteenwegen2022survey}
Vansteenwegen P, Melis L, Akta{\c{s}} D, Montenegro BDG, Vieira FS,
  S{\"o}rensen K (2022) A survey on demand-responsive public bus systems.
  Transportation Research Part C: Emerging Technologies 137:103,573

\bibitem[{Viergutz and Schmidt(2019)}]{viergutz2019demand}
Viergutz K, Schmidt C (2019) Demand responsive-vs. conventional public
  transportation: A matsim study about the rural town of colditz, germany.
  Procedia Computer Science 151:69--76

\bibitem[{Vinod and Sharma(2021)}]{vinod2021covid}
Vinod PP, Sharma D (2021) Covid-19 impact on the sharing economy post-pandemic.
  Australasian Accounting, Business and Finance Journal 15(1):37--50

\bibitem[{Wu and MacKenzie(2022)}]{wu2022evolution}
Wu X, MacKenzie D (2022) The evolution, usage and trip patterns of taxis \&
  ridesourcing services: evidence from 2001, 2009 \& 2017 us national household
  travel survey. Transportation 49(1):293--311

\bibitem[{Yao and Bekhor(2021)}]{yao2021dynamic}
Yao R, Bekhor S (2021) A dynamic tree algorithm for peer-to-peer ridesharing
  matching. Networks and Spatial Economics 21(4):801--837

\bibitem[{Yoon et~al(2022)Yoon, Chow, and Rath}]{yoonsimulation}
Yoon G, Chow JY, Rath S (2022) A simulation sandbox to compare fixed-route,
  semi-flexible transit, and on-demand microtransit system designs. KSCE
  Journal of Civil Engineering pp 1--20

\bibitem[{Young et~al(2020)Young, Farber, and Palm}]{young2020true}
Young M, Farber S, Palm M (2020) The true cost of sharing: A detour penalty
  analysis between uberpool and uberx trips in toronto. Transportation Research
  Part D: Transport and Environment 87:102,540

\bibitem[{Zhao et~al(2018)Zhao, Yin, An, Wang, and Feng}]{zhao2018ridesharing}
Zhao M, Yin J, An S, Wang J, Feng D (2018) Ridesharing problem with flexible
  pickup and delivery locations for app-based transportation service:
  Mathematical modeling and decomposition methods. Journal of Advanced
  Transportation 2018

\bibitem[{Zwick and Axhausen(2020)}]{zwick2020analysis}
Zwick F, Axhausen KW (2020) Analysis of ridepooling strategies with matsim. In:
  20th Swiss Transport Research Conference (STRC 2020)(virtual), IVT, ETH
  Zurich

\bibitem[{Zwick and Axhausen(2022)}]{zwick2022ride}
Zwick F, Axhausen KW (2022) Ride-pooling demand prediction: A spatiotemporal
  assessment in germany. Journal of Transport Geography 100:103,307

\end{thebibliography}

\end{document}